\documentclass[superscriptaddress,twocolumn,showpacs,preprintnumbers,prb]{revtex4}
\usepackage{graphicx}
\usepackage{times}
\usepackage{epsfig}

\begin{document}
\title{Unexpected Conductance Dip in the Kondo Regime  \\
of Linear Arrays of Quantum Dots}
\author{C.A. B\"usser, Adriana Moreo and Elbio Dagotto}
\affiliation{National High Magnetic Field Lab and Department of Physics,
Florida State University, Tallahassee, FL 32306}
\date{\today}

\begin{abstract}
Using exact-diagonalization of small clusters and Dyson equation
embedding techniques, the conductance $G$
of linear arrays of quantum dots is investigated. The Hubbard interaction
induces Kondo peaks at low temperatures for an odd number of dots.
Remarkably, the Kondo peak is split in half
by a deep minimum, and the conductance
vanishes at one value of the gate voltage. Tentative explanations for
this unusual effect are proposed, including an
interference process between two channels contributing to $G$, with one more
and one less particle than the exactly-solved
cluster ground-state. The Hubbard interaction
and fermionic statistics of electrons also appear to be 
important to understand this phenomenon. Although most of the calculations
used a particle-hole symmetric Hamiltonian and formalism, 
results also presented
here show that the conductance dip exists even when this symmetry is
broken. The conductance
cancellation effect obtained using numerical techniques
is potentially interesting, and other many-body techniques should be used to
confirm its existence.
\end{abstract}


\maketitle



\section{Introduction}

The possibility of destructive interference between
two or more wave functions is among the most remarkable phenomena
predicted by quantum mechanics. The effect can be observed when
electronic beams are split and then brought together after traveling 
paths of different lengths, or in a Aharonov-Bohm (AB) geometry -- such
as a ring -- where two equal-length paths nevertheless can carry
different phase factors in the presence of a magnetic flux. 
Recent advances in nanotechnology have made possible the
fabrication of quantum dots \cite{kastner} -- analogous to artificial 
atoms or molecules -- where these effects can be tested.
In fact, the AB effect was recently observed using a quantum
dot embedded in a ring in the Coulomb blockade regime \cite{AB}.
Another example of conductance cancellations are the
well-known Fano resonances\cite{fano} that occur
when charge can circulate through two paths:
one with a discrete level and the other with a continuum of states.
Many physical realizations of Fano resonances are known. For example,
when an atom is deposited on a metallic surface,
a scanning tunneling microscope (STM) tip probes transmission to the tip either
through the atom or directly from the surface, leading to current
cancellations \cite{tip}. In addition, theoretical studies predict that
Fano resonances should appear in  ``T-shaped'' geometries 
where an active dot -- connected to
left and right electrodes -- is also side-connected 
to another dot \cite{selman1}. 
A similar conductance cancellation
has been predicted using double quantum-dot
molecules attached to leads \cite{claro} and in a
2$\times$2 quantum-dot array \cite{claro2}. All these cancellations 
are caused by destructive interference among two different 
paths between conductors. Related cases
correspond to multiple-level dots with noninteracting 
electrons \cite{selman2}, which can also be 
rephrased as a many-dot problem connected to leads in ring-like geometries, 
leading naturally to conductance cancellations. 

\begin{figure}
\begin{center}
\epsfig{file=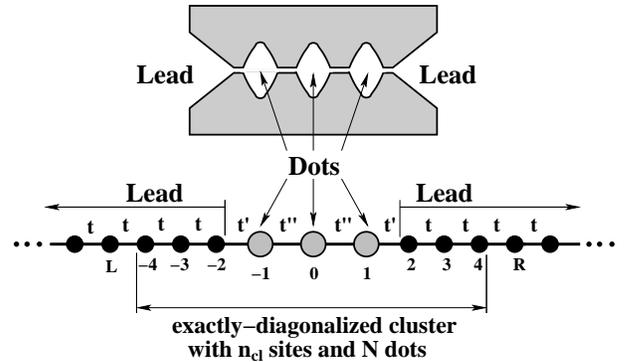,width=8.0cm}
\end{center}
\caption{Schematic geometry and hopping amplitudes
of the quantum-dot linear array studied here. The exactly-solved
cluster -- with $n_{\rm cl}$=9 sites and $N$=3 dots in this example -- 
includes some sites of the leads.
}
\label{Fig1}
\end{figure}

It is the purpose of this paper to report an unexpected 
conductance cancellation found in linear arrays of $N$ quantum
dots ($N$ odd and keeping only 
one level per dot). These arrays do not have any obvious 
real-space paths 
that may lead to explicit AB or Fano interferences to rationalize the
conductance zeros found here -- all electrons travel through the same
chain -- and in this respect the interference is exotic. 
In addition,
the reported cancellations occur only in the presence
of Coulomb interactions, and in the previously believed to be well-understood
Kondo regime \cite{glazman}. In fact, {\it the Kondo peaks are here found to be
split in half}, a curious effect that may be observable experimentally
at very low temperatures.
Although transport through many quantum dots\cite{coupled} 
or STM-engineered atomic
systems\cite{crommie} has received
considerable attention in recent years, experiments
using 3 dots or atoms have not been sufficiently accurate to address the
effect found in this paper.

Note that conductance dips have also been observed experimentally 
and theoretically
at high magnetic fields in a two-level single dot \cite{S-T1,S-T2}. 
This effect was ascribed to transitions between total spin S=1 and S=0 
dot states, a situation that does not seem to apply to our case with
an odd number of electrons in the ground state \cite{oguri}.

The paper is organized as follows. In Sec.\ref{sec:model}, the 
Hamiltonian and many-body technique are described. In Sec. \ref{sec:results}
the main results are presented, with emphasis on the conductance
cancellation for an odd number of dots. The case of an even number of
dots is also described, and in this situation there is no cancellation.
The dependence of the results with parameters in the model is presented
in this section as well. 
In Sec.\ref{sec:1dot}, results for 1, 4, and 5 dots
are briefly described. In Sec.\ref{sec:explanations}, possible explanations
of the conductance 
dip effect are presented. They include interference between conduction
processes  with one more and one less particle in the cluster, as well
as mappings into systems with the T-shape geometry that are known
to lead to interference. 
In Sec.\ref{sec:conclusions}, conclusions are
presented. 


Throughout the paper it is emphasized that confirmation 
of our results using other
techniques is important. Although the numerical 
studies presented below do not seem to be severely affected by size effects,
size dependences are sometimes very subtle. Thus, further work is 
needed to confirm the exotic conductance dip found here numerically.
If this confirmation occurs, the effect unveiled
in the present investigations -- an unexpected quantum interference process
in linear chains of dots -- should be searched for experimentally. The 
effort should be carried out at
sufficiently low temperatures such that  the dip structure becomes
visible.

\section{Model and Technique} \label{sec:model}

Keeping one level per dot, the Hamiltonian for our $N$-dot system coupled
to leads is
$H$=$H_{\rm dots}$+$H_{\rm leads}$+$H_{\rm int}$, where
\begin{equation}
H_{\rm dots} =  -t^{''} \sum_{i\sigma}(c^\dagger_{i\sigma} c_{i+1\sigma} 
+ h.c.)+  U\sum_{i} n_{i\uparrow} n_{i\downarrow},
\label{Eq1}
\end{equation}
\noindent represents the electronic hopping and Hubbard interaction
in the dots subsystem ($i$ labels the dots) using a standard notation. 
A gate voltage 
$V_{\rm g}\sum_{i\sigma} n_{i\sigma}$ of equal strength for 
the $N$ dots is also included. The term $H_{\rm leads}$ represents
the non-interacting electrons in the leads, with a nearest-neighbors
hopping amplitude $t$, while $H_{\rm int}$ is the hopping from the
ideal leads to the dots and its amplitude is $t^{'}$. Figure \ref{Fig1}
illustrates the geometry used in the study and conventions followed.
The Hamiltonian discussed here becomes particle-hole symmetric for 
the case $V_{\rm g}$=$-U/2$, precisely the gate voltage needed for
the conductance cancellation reported below. However, 
in subsection \ref{subsec:symmetries} other less symmetric models
were studied as well, and the zero in the conductance survives. Thus,
the dip reported in this paper does not seem to originate from a highly
symmetric Hamiltonian but its origin is more robust.

The zero-temperature, $T$=0, Green function $G_{\rm LR}(\omega)$ to
transfer charge from sites L to R (Fig.\ref{Fig1}) can be obtained by an
exact-diagonalization (Lanczos)
solution\cite{Elbio} of a cluster with $n_{\rm cl}$ sites
containing the $N$ dots. The exact
information about the cluster under study is supplemented by an
embedding procedure between the leads, already 
discussed in previous literature \cite{meth1,meth2}.
To reproduce the one-dot Kondo effect it is crucial that the
exactly-solved cluster contains also a small
portion of the lead\cite{meth2}, assumed also 
in a linear arrangement for
simplicity. The cluster size $n_{\rm cl}$ is chosen such that
$n_{\rm cl}$=$N$+2$n_{\rm odd}$, where $n_{\rm odd}$=1,3,5,...
As discussed before \cite{meth2},
with this convention the portion of the leads in the cluster contains
a zero-energy state that induces the Kondo effect already at the cluster
level, reducing finite-size effects.
The rest of the contacts is
incorporated using the Dyson equation 
$\hat{G}$=$\hat{g} + \hat{g} ~\hat{t} ~\hat{G}$,
where $\hat{g}$ is the exactly-known
Green function matrix of the cluster, $\hat{G}$
is the dressed Green-function matrix across the cluster from L to R,
and $\hat{t}$ is the matrix of hopping elements
connecting the cluster and leads. 
In the present study, the total
$z$-component of the spin is either 1/2 or -1/2 
for an odd number of sites in
the exactly-solved cluster. To respect particle-hole symmetry at
every step in the calculation, the cluster ground-state is here
taken as the sum (divided by $\sqrt{2}$) of the ground states of the
subspaces with total spin $z$-component 1/2 and -1/2.
This leads to Green functions for the ``up'' and ``down'' spins
that are identical.
Other conventions, discussed in the Appendix, lead to qualitatively
similar results regarding the presence of internal structure in 
the conductance Kondo peak.

To consider charge fluctuations, the cluster Green functions 
$\hat{g}_m$ for $m$ and $m$+1
electrons are combined. The mixed Green function $\hat{g}$ 
is written as $\hat{g}$=$(1-p) \hat{g}_m + p~\hat{g}_{\rm m+1}$. 
With the dressed Green function $\hat{G}$ from the Dyson eq.,
the total cluster charge is obtained
as $Q$=$-1/\pi \int_{-\infty}^{E_{\rm F}} 
\sum_j \mbox{Im} G_{jj}(\omega)$ (the sum in $j$ runs over the
cluster sites and $E_{\rm F}$ is the Fermi energy, assumed 0 in the
numerical calculations discussed below). 
On the other hand, the charge at the cluster 
in the mixed ($m$/$m$+1) state is
$q$=$(1$-$p)m$+$p(m$+$1)$ and, then, $p$ can be 
found self-consistently to satisfy 
$Q$=$q$ (in the region emphasized in the next section, with a $G$ cancellation, 
$q$$\sim$$n_{\rm cl}$).
Finally, using the Keldish formalism 
the conductance $G$ is written as \cite{meth2} 
${G}$=$(e/h)^2 t^2 |G_{\rm LR}(E_{\rm F})|^2 [\rho_{\rm leads}(E_{\rm F})]^2$.
The leads density-of-states (DOS) is $\rho_{\rm leads}(\omega)$, assumed here 
to be a semi-circle from -2$t$ to 2$t$ (the results are weakly 
dependent on this assumption).

Previous studies showed that this formalism -- combination 
of exact-diagonalization and embedding -- is sufficiently powerful to
reproduce the Kondo effect in electronic transport 
across one dot \cite{meth2}, and for this reason there is no {\it a priori}
reason why it would fail for more dots. Nevertheless, to be cautious
in our discussion below statements are included to alert the reader on
possible subtle size effects that could alter our conclusions.
As already mentioned, 
in the Appendix alternative conventions to our choice of spin
quantum numbers for the exactly-solved cluster ground state are also
discussed. These alternative conventions also lead to dips in the conductance
Kondo peak, as in the results presented in the following sections. 
Although more work is needed to confirm the existence of the
conductance dip reported below, 
the numerical results appear to be sufficiently robust that the effect
could even be observed experimentally at very low temperatures.

\begin{figure}
\begin{center}
\epsfig{file=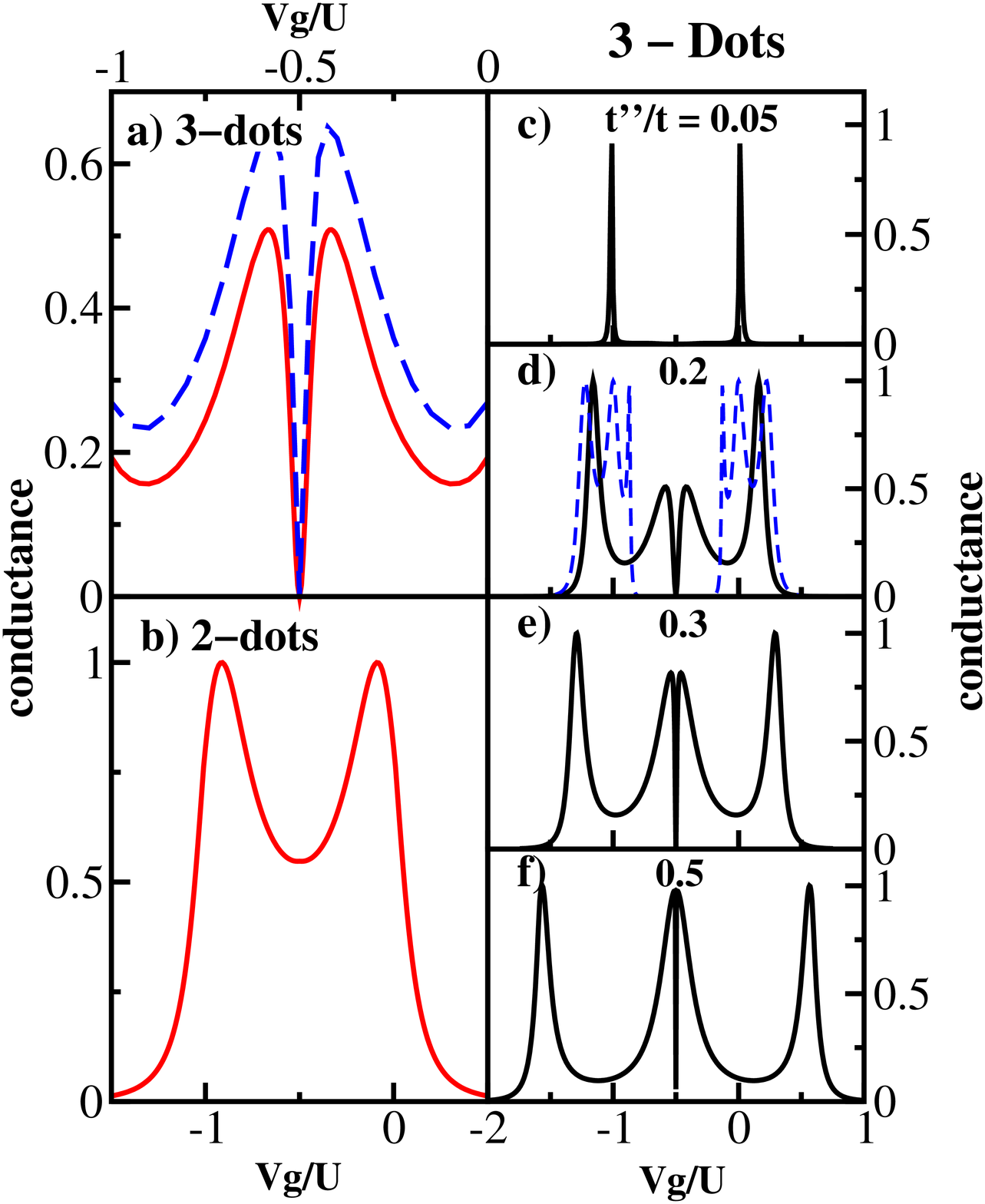,width=8.0cm}
\end{center}
\caption{Conductance (in units of $e^2/h$) across an array of quantum dots
vs. $V_{\rm g}/U$,  illustrating
the cancellation reported in this paper. The couplings are
$U/t$=1 and $t^{'}/t$=0.3 (a) corresponds to $N$=3 dots,
at $t^{''}/t$=0.2. 
The solid (dashed) lines corresponds to an exactly-solved
cluster of $n_{\rm cl}$=5 (9) sites. The size dependence suggests
that this effect will survive the bulk limit.
(b) is for 2 dots ($n_{\rm cl}$=4), 
same $t^{''}/t$ as in (a). (c-f) are results
in a wider range of $V_{\rm g}/U$ and varying $t^{''}$. For 
very small $t^{''}$, (c), the central dot is virtually decoupled and
no Kondo effect is observed in the scale used. With increasing
$t^{''}$ a central peak is found, always split as in (a).
Applying the method outlined in the text to a cluster that only 
has the dots (no extra lead sites), and then
incorporating the effect of the leads through the Dyson equation, 
the Kondo peak is effectively eliminated (the exactly-solved cluster
does not have states near the leads Fermi energy). By this
procedure, just the Coulomb blockade peaks are found, roughly
representing the high-temperature solution of the problem. This result
is shown with dashed lines in (d) for completeness.
}
\label{Fig2}
\end{figure}

\section{Results} \label{sec:results}

\subsection{Conductance Dip for an Odd Number of Dots}

The technique described in the previous paragraph was applied here to
the case of $N$$>$1 quantum dots forming a linear array. Our
original motivation was the study of transport in the regime of
large $t^{''}$ where $N$ odd (even) would lead to a quantum-dot
subsystem with spin 1/2 (0) and, as a consequence, the presence (absence) of
the Kondo effect as indeed occurs. However, studies at
intermediate couplings and hoppings regimes led to surprises. 
The most unexpected
result of the present effort is shown in Fig.~\ref{Fig2}a where the
conductance across 3 dots is shown for $V_{\rm g}$ near $-U/2$ (inducing
one electron per dot), at relatively small $t^{''}$. 
The shape of the broad peak (without the dip) resembles previous
Kondo-like results for one dot\cite{meth2}. Following standard 
arguments, this Kondo effect is obtained 
when the state of $N$ (odd) electrons carrying a net spin couples 
to the leads \cite{glazman,meth2}.
However, the peak is found to be
split in half by an unexpected 
zero in the conductance at exactly $V_{\rm g}$=$-U/2$. 
This cancellation is absent at $U$=0, where
$G$/$(e^2/h)$=1 at $V_{\rm g}$=0 since the exactly-solved cluster has a 
zero-energy state aligned with the leads Fermi-energy (assumed at 0). 
As $U$ increases,
Coulomb blockade and Kondo peaks are generated, but
the latter is always split 
by a zero at $V_{\rm g}$=$-U/2$. As a consequence, the effect 
appears to
originate in correlation effects induced by a nonzero $U$. 
Our study for
increasing $N$ suggests that the effect is present for any odd $N$, while
for $N$ even (2 as example, Fig.~\ref{Fig2}b) there is no conductance
cancellation
(the two peaks in Fig.~\ref{Fig2}b 
are related to the Kondo splitting -- without a zero -- of double
quantum dots, previously discussed \cite{meth2,split}).
The $t^{''}$-dependence shown in Fig.~\ref{Fig2}c-f suggests
that there is an intermediate hopping range where the zero conductance
effect could be observed\cite{comment3}, while 
both at very small and large $t^{''}$, its experimental observation will be
difficult.

The conductance cancellation Fig.~\ref{Fig2}a is unexpected
since there are no obvious real-space multiple paths that can
lead to interference. Electrons here travel through a one
dimensional geometry. Note also that one level per dot is kept 
in our analysis, and cancellations as in Ref.\onlinecite{selman2} are not
obviously present.

\begin{figure}
\begin{center}
\epsfig{file=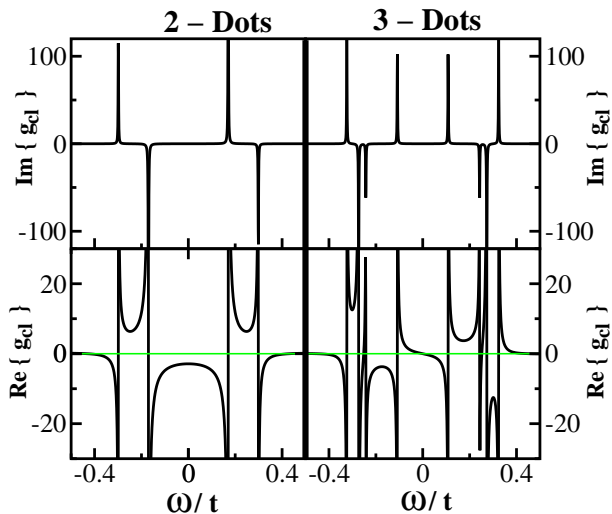,width=8.0cm}
\end{center}
\caption{Real and imaginary parts of the
cluster Green function
$g_{\rm cl}$ 
(from the first to the last site of the cluster) used to calculate $G$
through the Dyson equation. $\omega$ is in units of $t$.
Shown are exact results for $N$=2 dots ($n_{\rm cl}$=4 cluster) 
and $N$=3 dots ($n_{\rm cl}$=5 
cluster). The couplings are as in Fig.~\ref{Fig2}a. Note that
for 3 dots the real part vanishes at $\omega$=0, while for 2 dots
it is finite. The different behavior under $\omega$$\rightarrow$$-\omega$
for odd and even number of dots causes the cancellation of the conductance
of the former, discussed in the text.}
\label{Fig3}
\end{figure}

To gain insight on the origin of
the reported  
phenomenon, Fig.~\ref{Fig3} shows the real (Re) and imaginary (Im)
components of the {\it cluster} Green function from one extreme
of the cluster to the other (denoted
$g_{\rm cl}$), for 2 and 3 dots. The overall
conductance emerges 
from the behavior of $g_{\rm cl}$ at $\omega$=0, in the embedding procedure. 
Clearly, the results for 2 and 3 dots
have different symmetry properties under $\omega$$\rightarrow$$-\omega$:
while for 2 dots Re($g_{\rm cl}$) is even, for 3 dots it is odd generating
a zero at $\omega$=0. Since both imaginary parts cancel at $\omega$=0,
then Re($g_{\rm cl}$) and Im($g_{\rm cl}$)
are zero (Re nonzero) for odd (even) number of dots
(this rule was verified numerically beyond the 2- and 3-dots example shown). 
If $g_{\rm cl}$=0, the Dyson-equation embedding procedure cannot generate
a nonzero conductance.
If $U$$\rightarrow$0, the two peaks closest to $\omega$=0 
in Fig.~\ref{Fig3} (right upper panel) 
merge and the cancellation does not occur.

\subsection{Analysis of Size Effects}

\begin{figure}
\begin{center}
\epsfig{file=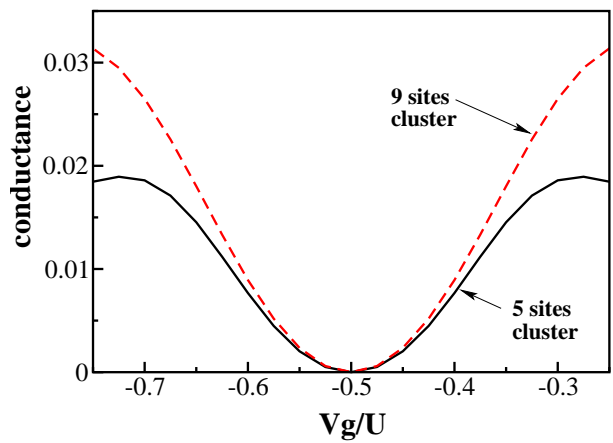,width=8.0cm}
\end{center}
\caption{
Conductance at $U$=1, $t$=1, and $t^{'}$/$t$=0.3, as used in
Fig.\ref{Fig2}a, and for $t^{''}$/$t$=0.075. Results for 3 dots and 
cluster sizes $n_{\rm cl}$ of 5 and 9 sites are indicated. The maximum in the
conductance does not seem to move toward $V_{\rm g}$/$U$=-0.5 as $n_{\rm cl}$
grows, suggesting that the dip will survive the bulk limit.
}
\label{Fig7}
\end{figure}

An important aspect of the methodology discussed here, and in
previous literature, involves the exact solution of a cluster
followed by an embedding procedure. From the cluster size
dependence it is possible to infer whether a particular feature
under study will survive the bulk limit or not. Unfortunately, the
CPU time rapidly grows with the cluster size since the cluster 
Green functions at all distances are needed for the Dyson equations,
and each Green function is calculated with approximately 
one hundred steps in the continued-fraction procedure \cite{Elbio}.
This limits our detailed study of the conductance dip to clusters with
5 and 9 sites (while a few values of the gate voltage can still be investigated
using 13 sites). In Fig.\ref{Fig7}, results for $n_{\rm cl}$=5 and 9
are presented at a small value of $t^{''}$. This small hopping was used
to amplify the region where the dip dominates (as $t^{''}$/$t$ increases,
the dip width is reduced as shown in Fig.\ref{Fig2}c-f). The results in
the figure show that the dip {\it survives} the increase of the cluster size,
and the maximum in the conductance actually is located even further away
from $V_{\rm g}$/$U$=-0.5 as the $n_{\rm cl}$ grows. While this is not
a definite proof, it is strongly suggestive that the conductance dip is
not an artifact of the many-body procedure and cluster-size used, but
it may be a real effect present in the bulk. Nevertheless, it is desirable
to have independent tests of our results using other many-body methods
to fully confirm our conclusions. 

\subsection{Survival of the Dip Reducing the Symmetries of the Hamiltonian}
\label{subsec:symmetries}

In real quantums dots, the electron-hole symmetry of the Hamiltonian
used in previous sections cannot be achieved
since the Fermi level lies just a small fraction of eV 
above the bottom of the conduction band. 
However, it is always possible to find a gate potential where the main 
levels involved -- $V_{\rm g}$ and $V_{\rm g}$+$U$ -- are symmetrically 
located around $E_F$.
Since all the energy scales important for the Kondo effect ($U$ and
$t^{'}$) are of the order of meV \cite{Golhaber2} these levels are
very close of $E_F$, achieving an approximate 
electron-hole symmetry.

\begin{figure}
\begin{center}
\epsfig{file=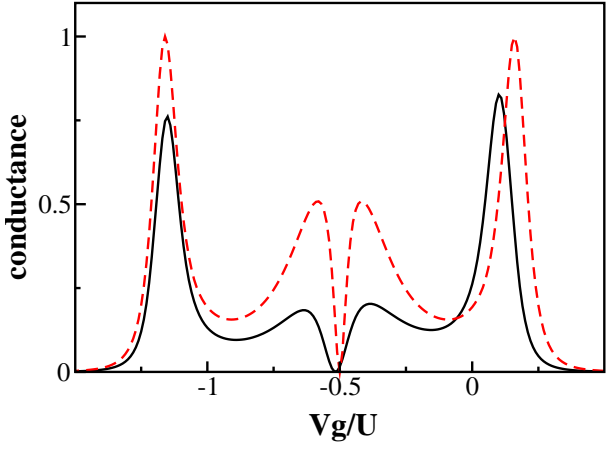,width=8.0cm}
\end{center}
\caption{Conductance for the 3-dot system, with 5 sites in the
exactly-solved cluster (solid line). In this figure
the on-site Hubbard $U$ couplings, as well
as the on-site energies $\epsilon$, of the three interacting sites
are given random values of amplitude 0.01$t$ in addition
to the uniform values used in Fig.\ref{Fig2} (namely $U$=1 and $\epsilon$=0).
The hopping amplitudes $t^{'}$ and $t^{''}$ are also varied
as follows: from the left lead to the first dot  $t^{'}$ is 0.3 (in 
units of $t$), the next hopping amplitude is $t^{''}$=0.2, then
$t^{''}$=0.12, and finally $t^{'}$=0.35 for the connection between the 
last dot and the
right lead. It is observed that the vanishing of the conductance still
occurs although with a small shift in its position.
For comparison, the result of Fig.\ref{Fig2}a in the particle-hole symmetric
case is also shown (dashed line).
}
\label{Fig8}
\end{figure}

In addition to the previous argument, it is possible to repeat the
calculations presented before for cases where the particle-hole
symmetry is not present. This can be achieved, for instance, by merely 
adding small random components to the on-site Hubbard energies
and, in addition, introducing small site energies. To further break symmetries
of the problem, the hopping $t^{'}$ from the 3-dot region to the rest can also
be made different on the right and the left, and even the two internal
dot-dot hopping amplitudes $t^{''}$ can be made different as well. 
One representative result of this study
is shown in Fig.\ref{Fig8}. It is interesting to observe that the
dip in the conductance survives the breaking of symmetries in the model.
As a consequence, the effect appears to be robust and independent of fine
details in the analysis.

\section{Results for Number of Dots 1, 4, and 5} \label{sec:1dot}

\begin{figure}
\begin{center}
\epsfig{file=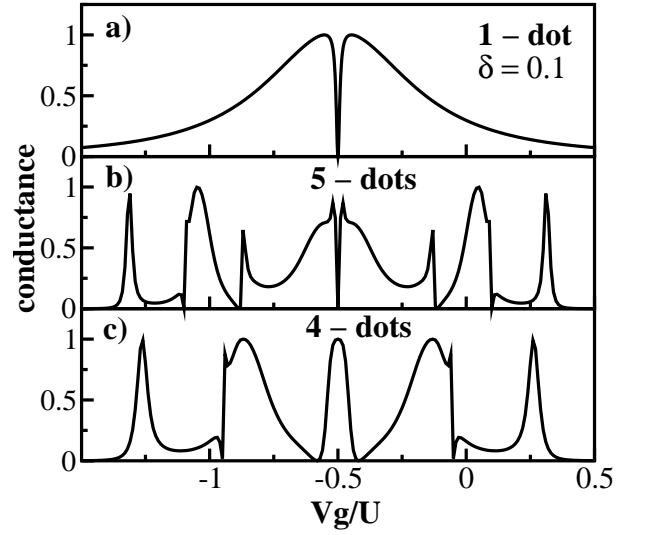,width=8.0cm}
\end{center}
\caption{
(a) $G$/$(e^2/h)$ vs. $V_{\rm g}/U$ for
1 dot, $t^{'}/t$=0.3, introducing on-site energies
$\pm\delta$=0.1 (see text). (b-c) are results for
 clusters with 5 and 4 dots, showing a
rich structure. Parameters are as in Fig.~\ref{Fig2}a. 
}
\label{Fig4}
\end{figure}

The case of one dot is special. 
If a ${n_{\rm cl}}$ cluster with an odd number of sites
(e.g., +-o-+) is solved exactly, the degeneracy
between ${n_{\rm cl}}$-1, ${n_{\rm cl}}$, and ${n_{\rm cl}}$+1 
{\it remains} even for nonzero $U$.
This is an accidental degeneracy that avoids the dip splitting of the
Kondo peak found with 3 or more (odd) dots. However, by simply adding
on-site energies $\delta$ and $-\delta$ at the first and last sites
of the cluster, the accidental degeneracy at $U$$\neq$0 
is removed and now a conductance cancellation 
occurs as in the other cases (Fig.~\ref{Fig4}a) \cite{comment4}. The
dip phenomenon appears to be general and robust.

Our study also extended beyond the 2- and 3-dot cases. For example,
Fig.~\ref{Fig4}b illustrates results obtained for 5 dots (7-site cluster).
Here, once again, $G$=0 at $V_{\rm g}$=$-U/2$. In addition, a rich
structure if observed at higher frequencies with multiple conductance
cancellations that resemble Fano resonances. Their origin is similar
to those discussed for 3 dots and emerge from cancellations between
``competing'' poles at close distance in the Green function. Even
4 dots (Fig.~\ref{Fig4}c) shows a highly nontrivial structure, also
with cancellations away from $V_{\rm g}$=$-U/2$. The
richness unveiled in the conductance properties of linear-dot chains
once unbiased accurate many-body techniques are used is 
remarkable.


\section{Possible Explanations of Conductance Cancellation} \label{sec:explanations}

\subsection{Interference Between States with ${n_{\rm cl}\pm 1}$ particles}

The results of the previous paragraph suggest that simple
symmetry arguments involving just two states of the
entire Hilbert space -- those closest to $\omega$=0 -- should
be sufficient to understand the effect. With this in mind, consider
the $n_{\rm cl}$-site-cluster Green-function of interest expanded
in the basis of Hamiltonian eigenstates as 
\begin{eqnarray}
 &  g_{\rm cl}(\omega) =    \sum_l 
{{\langle 0,{n_{\rm cl}} |c_1|l,{n_{\rm cl}}+1\rangle   
\langle l,{n_{\rm cl}}+1 |c^\dagger_{n_{\rm cl}}|0,{n_{\rm cl}}\rangle}\over{\omega + E_l -E_0 +i\epsilon}} 
\nonumber \\
 &+  \sum_m{{\langle 0,{n_{\rm cl}} |c^\dagger_{n_{\rm cl}}|m,{n_{\rm cl}}-1\rangle 
\langle m,{n_{\rm cl}}-1 |c_1|0,{n_{\rm cl}}\rangle}\over{\omega + E_0 -E_m +i\epsilon}},
\label{Eq2}
\end{eqnarray}
where $\epsilon$$\rightarrow$0 (10$^{-7}$ in practice, and 10$^{-2}$ for
the DOS). 
In $|j,n\rangle$, $j$ labels states
of $n$ particles.
The Hubbard Hamiltonian is particle-hole (p-h) symmetric if
$V_{\rm g}$=$-U/2$. The explicit p-h transformation is
$c^\dagger_{i\sigma}$$\rightarrow$$(-1)^{i}{c}_{i\sigma}$.
The empty state is mapped into the fully occupied state. 
It can be shown that 
$\langle 0,{n_{\rm cl}} |c_1|l,{n_{\rm cl}}+1\rangle$$\rightarrow$$(-1)^p \langle m,{n_{\rm cl}}-1 |c_1|0,{n_{\rm cl}}\rangle$
for ${n_{\rm cl}}$=$2p$+$1$ or ${n_{\rm cl}}$=2$p$, with $p$=integer. In addition,
$\langle l,{n_{\rm cl}}+1 |c^\dagger_{n_{\rm cl}}|0,{n_{\rm cl}}\rangle$$\rightarrow$ 
$(-1)^{r}\langle 0,{n_{\rm cl}} |c^\dagger_{n_{\rm cl}}|m,{n_{\rm cl}}-1\rangle$,
where $r$=$p$ if ${n_{\rm cl}}$=$2p$+$1$, and $r$=$p$+$1$ if ${n_{\rm cl}}$=$2p$.
Isolating a pair of states $|l,{n_{\rm cl}}+1\rangle$ and $|m,{n_{\rm cl}}-1\rangle$
connected by the p-h transformation -- namely with equal energies
relative to $E_0$ --
this leads to a simple contribution to $g_{\rm cl}(\omega)$
of the form 
$[{{AB}/(\omega+\omega_l+i\epsilon)}] \pm
 [{{AB}/(\omega-\omega_l+i\epsilon)}]$,
where $A,B$ are numbers and $\omega_l$=$E_l$-$E_0$, assumed nonzero. 
The $+$ $(-)$
sign corresponds to an {\it odd} ({\it even})
number ${n_{\rm cl}}$ of cluster sites. Clearly, for ${n_{\rm cl}}$=odd, 
$Re[g_{\rm cl}(\omega)]$ is odd under
$\omega$$\rightarrow$$-\omega$, and, thus, it cancels at $\omega$=0.
For the other case, ${n_{\rm cl}}$=even, there is no cancellation since
the real part is even. Since the embedding process cannot generate
a nonzero conductance if the cluster has a vanishing Green function,
then the overall conductance is zero 
for $V_{\rm g}$=$-U/2$ and ${n_{\rm cl}}$ odd.

The proof of the results of the previous paragraph has been 
mainly computational,
using the entire Hilbert space for small values of ${n_{\rm cl}}$. However,
a simpler qualitative understanding 
can be obtained for example considering ${n_{\rm cl}}$=3 and using 
($|\uparrow \downarrow \uparrow \rangle$+
 $|\downarrow \uparrow \downarrow \rangle$)/$\sqrt{2}$ as a
simplified ${n_{\rm cl}}$-particle 
ground-state $|0,{n_{\rm cl}}\rangle$. For $| l,{n_{\rm cl}}+1\rangle$,
$c^\dagger_{-1,\uparrow}$$|0,{n_{\rm cl}}\rangle$ can be used,
and  $c_{-1,\downarrow}$$|0,{n_{\rm cl}}\rangle$
for $| l,{n_{\rm cl}}-1\rangle$.
For these simplified states, it can be easily shown
that $|l,{n_{\rm cl}}+1\rangle$ transforms under 
p-h to $|l,{n_{\rm cl}}-1\rangle$ for ${n_{\rm cl}}$=odd,
and to $-|l,{n_{\rm cl}}-1\rangle$ for ${n_{\rm cl}}$=even. 
After simple algebra, recalling
that the matrix elements are real, and being careful with
the signs arising
from fermionic anticommutations, the p-h transforms of the matrix
elements are found,
completing the proof. This ${n_{\rm cl}}$=3 derivation can easily 
be extended to arbitrary ${n_{\rm cl}}$.


The previous explanation of the anomalous zero conductance
emphasizes the competition, and eventual
interference, between two states that contribute to the 
cluster Green function. The key aspect is the relative
sign of the matrix elements for the two poles, which leads
to interference for an odd number of cluster sites.
Let us discuss these aspects more intuitively,
and also explain why at $U$=0 the effect is not present.
Consider as example a 5-site cluster with 3 dots
(schematically +-o-o-o-+, o=dot, +=lead site). At $U$=0 this cluster, and any cluster with a total
number of sites odd and Hamiltonian Eq.~\ref{Eq1}, has a zero
energy eigenvalue. This implies a {\it degeneracy} between
the lowest-energy states with 4, 5, and 6 particles 
(or ${n_{\rm cl}}$-1, ${n_{\rm cl}}$, ${n_{\rm cl}}$+1 
particles for an ${n_{\rm cl}}$ (odd) cluster) since populating
the zero-energy state has no energy cost. With the Fermi energy
of the metal at 0 as well, there is a direct channel for conductance
through the dots and no cancellation. However, when $U$ is switched
on, the degeneracy is removed since there is a penalization
for having a site with two electrons (directly related to an empty site
by p-h symmetry). The doubly-occupied site, inevitable for ${n_{\rm cl}}$+1
electrons on ${n_{\rm cl}}$ sites, is located with more probability
outside the set of
$N$ dots, i.e. in the lead segments included in the exactly-solved cluster.
This produces a finite but small splitting $\Delta E$,
substantially smaller than $U$.
As $U$ increases, a Kondo peak is formed at $V_{\rm g}$=$-U/2$, as
previously discussed\cite{meth2}, but with a dip of width $\Delta E$ 
in the middle. Note that for ${n_{\rm cl}}$ even, there is no zero
in the cluster and no dip in the conductance. However, 
states approach zero energy as ${n_{\rm cl}}$ (even) 
increases and eventually
as ${n_{\rm cl}}$$\rightarrow$$\infty$ a common limit of zero conductance 
for both ${n_{\rm cl}}$ odd and even is expected.

\begin{figure}
\begin{center}
\epsfig{file=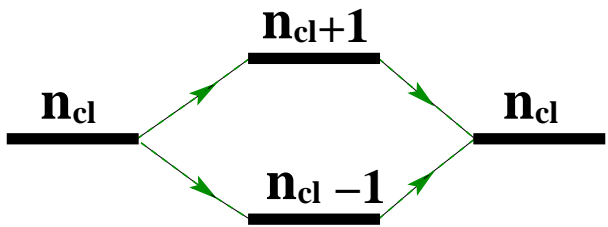,width=8.0cm}
\end{center}
\caption{The two ``paths'' that lead to the conductance cancellation
involve intermediate states of ${n_{\rm cl}}$+1 and ${n_{\rm cl}}$-1 electrons.
}
\label{Fig6b}
\end{figure}

The $G$ cancellation arises from the $U$-induced splitting of the
${n_{\rm cl}}$+1 and ${n_{\rm cl}}$-1 
states from the ${n_{\rm cl}}$ ground state. More intuitively,
for charge to transport through a cluster or molecule there are
two basic processes, that here interfere. In one case an electron
first jump to the cluster, leading to ${n_{\rm cl}}$+1 particles inside, and 
then an electron exits. In the other case, first an electron leaves the cluster
(i.e (${n_{\rm cl}}$-1) 
electrons in the cluster intermediate state), and then another
gets in. These two intermediate states corresponds to two ``paths''
in a quantum-mechanical formulation, and they do not need actual different
real-space trajectories to interfere with one another 
(Fig.~\ref{Fig6b}).

\subsection{Two-Paths Interference in a One-dimensional Multidot
System}

\begin{figure}[h]
\begin{center}
\epsfig{file=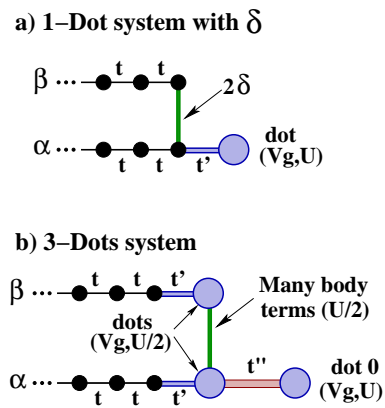,width=5.0cm}
\end{center}
\caption{Illustration of the transformation to the $\alpha$-$\beta$
basis described in the text. (a) corresponds to
1-dot with diagonal energies $\pm\delta$ in the 
sites next to the dot. (b) corresponds to 3-dots.
}
\label{Fig5f}
\end{figure}

An alternative explanation of the conductance-cancellation effect
described in this paper is the following.
The ${n_{\rm cl}}$-site cluster, with ${n_{\rm cl}}$ odd, 
has reflexion symmetry around 
the central dot (here denoted by 0). This suggests a change of
basis defined by 
$d_{\alpha i\sigma}$=$(c_{i\sigma}+c_{-i\sigma})/\sqrt{2}$ 
and 
$d_{\beta i\sigma}$=$(c_{i\sigma}-c_{-i\sigma})/\sqrt{2}$,
where the sites $i$=1,2,... (-1,-2,...) are on the right (left)
of the central dot, which is left invariant by this transformation.
It can be shown that for just one dot, $N=1$, the system in the 
new basis is equivalent to one-dot at the end of a semi-infinite
chain coupled to the $\alpha$ band, and decoupled from a $\beta$ band
(as sketched in Fig.~\ref{Fig5f}a, for $\delta$=0). 
This geometry corresponds to a
one-channel Kondo problem, with a concomitant peak in electronic
transport.
In this new basis, the Green function used to calculate the
conductance can be written as 
$G_{LR}(\omega)$=1/2($G_{\alpha\alpha}(\omega)$-$G_{\beta\beta}(\omega))$.
Then, an interference in the conductance occurs when
$G_{\alpha\alpha}(E_F)$=$G_{\beta\beta}(E_F)$.

As discussed in Sec.\ref{sec:1dot}, 
consider now a diagonal energy $\delta$ 
at site $i$=+1 and $-\delta$ at $i$=-1 
(i.e. immediately to the right and left of the active dot).
Transforming the operators, the previously decoupled 
bands $\alpha$ and $\beta$ 
are now effectively connected by a hopping term of strength $2\delta$ 
(Fig.~\ref{Fig5f}a). In this representation, it is easy to visualize
a possible interference between processes involving 
the direct hopping $\alpha$$\rightarrow$$\beta$ and those where jumping
to and from the dot is part of the path. This abstract-space representation
establishes a connection with the real-space $T$-geometry
interference previously discussed\cite{selman1}.

For $N$$>$1 the results are not as conclusive, but still suggestive.
Consider as an example $U$=0 and $N$=3. In this case, 
the two channels $\alpha$ and $\beta$, each supplemented by 
one dot, decouple from one another. The central dot is coupled to
the $\alpha$ channel (Fig.~\ref{Fig5f}b).
At $U$$\neq$0 a ``many-body'' coupling proportional to $U/2$ 
links the two channels. This
coupling -- too cumbersome to write here explicitly --
contains ``spin-flip'' contributions between $\alpha$ 
and $\beta$, density-density interactions, and even a two-electron
hopping term. The $T$-geometry appears once again, suggesting possible
interferences, but now with an 
$\alpha$$\rightarrow$$\beta$ effective ``hopping''
which is very complicated. Nevertheless, this is sufficiently illustrative
since for $N$ even none of the two channels $\alpha$-$\beta$ have
an extra dot attached, and the entire system is effectively
linear with no
obvious sources of interference, as indeed observed numerically.
Then, {\it $N$ odd and even are fundamentally different 
in this representation},
with the odd having possible sources of interference.

\section{Conclusions} \label{sec:conclusions}

Using numerical techniques, in this paper it has been argued  
that the conductance Kondo peak of an odd number
of quantum dots forming a linear array presents nontrivial internal structure in the form
of a dip. The calculation has passed many tests, but the authors
acknowledge that the reported result is quite unexpected and for
this reason other theoretical
techniques should be employed to test our predictions.
If the present results are confirmed in the near future,
the search for the ``Kondo dip'' in experiments should be carried out.
The experimental observation of the conductance dip
reported here may require considerable effort.
Realizations of the linear-array
geometry using atoms and employing STM  techniques to measure conductances 
are difficult. For instance,
atoms attract, and three of them on a metallic surface tend to form triangles
rather than chains \cite{crommie-private}. In addition, finite temperature
effects will likely tend to fill the dip in $G$, and 
temperatures even lower than
usually employed will be needed to see the effect. At present,
the characteristic energy regulating the width of the dip is still
unknown since the method used in the paper works only at zero temperature.
In spite of these caveats, the 
interference discussed here is sufficiently novel and interesting that its
experimental confirmation and theoretical extension to other types of
arrays should be actively pursued.

\section{acknowledgments}

This work was supported by the NSF grants DMR-0303348 and  DMR-0122523.
Conversations with Y. Meir, E. Anda, J. Verg\'es, G. Chiappe,
S. Ulloa, L. Glazman, and S. Hershfield are gratefully acknowledged.

\section {Appendix: Results using other conventions}

\begin{figure}[h]
\epsfig{file=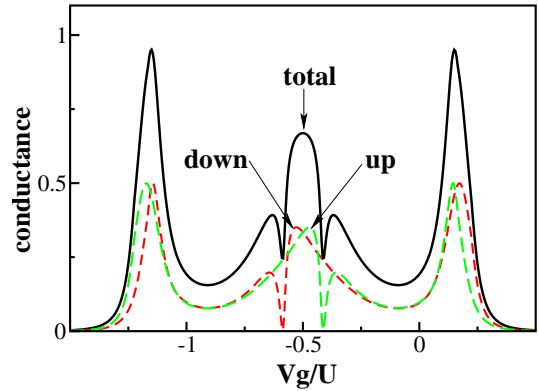,width=7.0cm}
\caption{
Conductance vs. gate voltage calculated using a 5 sites (3 dots)
cluster state
with total spin $z$-component equal to 1/2 (as opposed to the
equal-weight 1/2 and -1/2 used in the rest of the paper). Shown
with dashed lines are the dressed conductances (divided by 2) for the ``up''
and ``down'' channels, each showing a zero at values of the
gate voltage close to -0.5$U$. The solid line is the sum.
In the sum, two dips can be observed. Although the result is 
quantitatively different
using one convention or the other for the spin of the cluster state,
the fact that the Kondo peak has internal structure in the form of
dips is qualitatively the same in both cases.
}
\label{Fig9b}
\end{figure}

As discussed in Sec.\ref{sec:model}, the state representing the cluster
with an odd number of sites (and dots)
used in this effort is the equal-weight sum of the states with total
$z$-component of spin, $S^{\rm z}_{\rm tot}$, 
equal to 1/2 and -1/2. By this procedure the
particle-hole symmetry is respected at every step in the calculation.
However, other conventions have been used in recent literature.
For example, related work in Ref.~\onlinecite{willy} considers only
the cluster state with $S^{\rm z}_{\rm tot}$=1/2. With this convention
the cluster Green functions for spins 1/2 and -1/2 are different
(although their sum is independent of the relative weights 
of the two states). If these
``up'' and ``down'' Green functions are independently
dressed through the Dyson formalism,
conductances for the ``up'' and ``down'' channels are obtained.
Individually, {\it each of these conductances contains a zero} quite similar
to the results shown in the bulk of the present paper. However, the location
of the zero is different for the two channels, 
slightly shifted left and right 
from $V_{\rm g}/U$=-0.5 (see Fig.\ref{Fig9b}) due to the breaking 
of the particle-hole symmetry. When the two conductances
are added (and dividing by 2), 
the zeros are no longer present since the ``up'' and ``down''
contributions have the cancellation at different gate voltages. However,
even accepting this symmetry-breaking alternative convention to carry out the 
calculation, two dips are clearly found in the overall result, as also
shown in Fig.\ref{Fig9b}. 
It is expected that the two procedures (with and without explicit particle-hole
symmetry in the calculation) will
lead to the same result in the bulk limit, and only further work
can clarify which convention is the best given the inevitable
size constraints of the present numerical 
technique.

\begin{figure}[h]
\epsfig{file=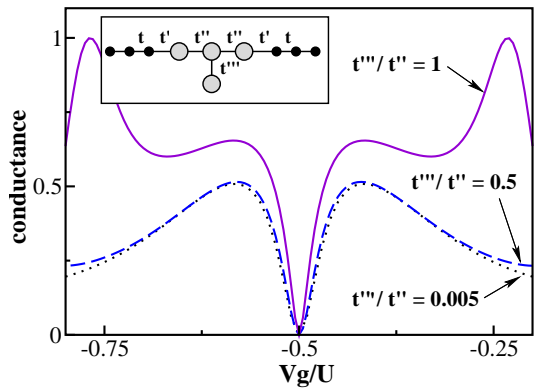,width=7.0cm}
\caption{
Conductance vs. gate potential for the $T$-geometry 
system (shown in the inset) proposed to avoid the ground-state
degeneracy at $V_{\rm g}/U$=$-0.5$. The parameters $U$, $t$, 
$t^{'}$, and $t^{''}$
are the same as in Fig.~\ref{Fig2}a. The point line corresponds
to results at $t^{'''}/t^{''}$=0.005,
the dashed line at $t^{'''}/t^{''}$=0.5, and the solid line  
at $t^{'''}/t^{''}$=1. The dip is present in all cases.
}
\label{Fig10}
\end{figure}

Finally, to avoid the cluster ground-state ambiguity 
problem an alternative geometry sketched in the
inset of Fig.\ref{Fig10} can be used. In this new system, one extra dot is laterally
coupled to the central dot with a hopping amplitude $t^{'''}$. 
Clearly, when $t^{'''} \to 0$ this system is equivalent to the 3 dots system 
described in the bulk of the paper. For $V_{\rm g}/U$=-0.5 
and one particle per site in the cluster, the
extra dot adds one electron to the system studied before
giving an $S^{\rm z}_{\rm tot}$=0 for the (nondegenerate)
ground state. This trick eliminates the ``up'' and ``down'' degeneracy
at all finite values of $t^{'''}$ (while at $t^{'''}$=0, the degeneracy is 
recovered). Repeating the calculation as in Fig.\ref{Fig2}a,  the
conductance for the new system once again presents a symmetric dip for all the 
values of $t^{'''}$ investigated, even including the very small $t^{'''}$ regime
where the extra dot and linear array are nearly decoupled. Then, once again
it is concluded that different procedures to carry out the
calculations lead all to the same qualitative conclusions. Both Figs.\ref{Fig9b} 
and \ref{Fig10} show that the main point of the present paper remains the same irrespective
of the convention: The Kondo peak of quantum-dot arrays with
an odd number of dots appears to have 
internal structure in the form of dips as the gate voltage is varied.

\begin{thebibliography}{99}
\bibitem{comment3} $t^{''}$ can be modified by controlling
the tunneling to and from the central dot using other appropriate voltages.
\end{thebibliography}
\end{document}